\documentclass[10pt]{article}
\usepackage[a4paper,margin=1.8cm]{geometry}
\usepackage[T1]{fontenc}
\usepackage[utf8]{inputenc}
\usepackage{graphicx}
\usepackage{multirow}
\usepackage{amsmath,amssymb,amsfonts}
\usepackage{amsthm}
\usepackage{mathrsfs}
\usepackage{xcolor}
\usepackage{textcomp}
\usepackage{manyfoot}
\usepackage{booktabs}
\usepackage{algorithm}
\usepackage{algorithmicx}
\usepackage{algpseudocode}
\usepackage{listings}

\graphicspath{{Figures/}{images/}}
\newcommand{\href}[2]{#2}
\newcommand{\url}[1]{\texttt{#1}}
\newcommand{\email}[1]{\texttt{#1}}
\newcommand{\backmatter}{}
\newcommand{\bmhead}[1]{\section*{#1}}

\begin{document}
\makeatletter
\begin{@twocolumnfalse}
\thispagestyle{plain}
\begin{center}
{\small\scshape Preprint\par}
\vspace{0.8em}
{\LARGE\bfseries Optical Readout of Reconfigurable Layered Magnetic Domain Structure in CrSBr\par}
\vspace{0.9em}
{\normalsize Aleksandra {\L}opion$^{1}$, Pierre-Maurice Piel$^{1}$, Thomas Kliewer$^{1}$, Manuel Terbeck$^{1}$, Jan-Hendrik Larusch$^{1}$, Jakob Henz$^{1}$, Marie-Christin Hei{\ss}enb{\"u}ttel$^{2}$, Kseniia Mosina$^{3}$, Thorsten Deilmann$^{2}$, Michael Rohlfing$^{2}$, Zdenek Sofer$^{3}$, Ursula Wurstbauer$^{1}$\par}
\vspace{0.8em}
{\footnotesize
$^{1}$ Institute of Physics and Center for Soft Nanoscience (SoN), University of M{\"u}nster, Wilhelm Klemm Str. 10, M{\"u}nster 48149, Germany\par
$^{2}$ Institute of Solid State Theory, University of M{\"u}nster, Wilhelm Klemm Str. 10, M{\"u}nster 48149, Germany\par
$^{3}$ University of Chemistry and Technology, Technicka 5, Prague 16628, Czech Republic\par}
\vspace{0.5em}
{\footnotesize\textit{Correspondence: } Aleksandra {\L}opion (\email{alopion@uni-muenster.de}) ~|~ Ursula Wurstbauer (\email{wurstbauer@uni-muenster.de})\par}
\vspace{0.8em}\hrule\vspace{0.8em}
\begin{minipage}{\textwidth}
\small\textbf{Abstract. }
The van der Waals magnetic semiconductor CrSBr combines multistable magnetic order with strong light--matter coupling, enabling optical access to a rich and reconfigurable layered magnetic domain structure. A purely optical, non-destructive, and non-contact readout of layered magnetic configurations is realized here by magneto-reflectance measurements and interpreted using an optical multilayer model. The magnetic state is tunable by applied magnetic fields and by interfacing CrSBr with the antiferromagnet MnPS$_3$. Applying an external magnetic field along the easy axis drives the hysteretic antiferromagnetic--to--ferromagnetic transition, which is not universally binary but instead develops through a cascade of intermediate magnetic configurations whose multiplicity and stability scale systematically with layer thickness and can be tailored by magnetic interfaces. The intertwined optical and magnetic properties of CrSBr provide a readout mechanism for information encoded in and processed through its magnetic configuration that is compatible with modern on- and off-chip photonic and electronic technologies. These properties identify CrSBr as a promising platform for intelligent matter and for spin-optoelectronics, in particular for neuromorphic architectures that can learn and evolve in response to changing environments.

\end{minipage}\par
\vspace{0.7em}
{\small\textbf{Keywords: } van der Waals magnets | CrSBr | magneto-optical readout | layered magnetic domains | intelligent matter  
\par}
\vspace{0.8em}\hrule
\end{center}
\vspace{1.2em}
\end{@twocolumnfalse}
\makeatother

\section{Introduction}
In magnetic systems, the combination of reconfigurability and hysteresis is particularly appealing because it enables both writing and storage of information. Antiferromagnets (AFM) offer ultrafast dynamics and multistability, whereas ferromagnets (FM) provide robust binary states but may suffer from magnetostatic crosstalk \cite{Jungwirth2016,Baltz2018,Zutic2004}. Optically accessible AFM-derived states therefore open routes towards robust memory functionalities, spin-driven optoelectronics, and neuromorphic architectures. More broadly, materials that can learn, process, store, and distribute information through interactions with their environment are discussed in the context of `intelligent matter' \cite{kaspar2021rise}. However, although multistability and memory are intrinsic to many magnetic systems with long-range order, direct and universal readout of reconfigurable magnetic states remains challenging.

In this context, 2D van der Waals (vdW) magnets provide an exceptional platform for reconfigurable long-range magnetic order. Their magnetic properties are governed by highly anisotropic intra- and interlayer exchange interactions, while weak interlayer vdW bonding enables efficient tuning of interlayer coupling, electronic structure, and magnetic order through stacking, electrostatic gating, strain, and interfacial engineering \cite{Huang2017,Gong2017,Gong2019Science2DMagnets,Blei2021Review,TabatabaVakili2024NatCommunCrSBrExcitonMagnetism,Bae2022,cenker2022reversible,deng2018gate,chen2019direct,song2021direct}. This unusual combination of magnetic tunability and structural flexibility makes 2D vdW magnets highly attractive for next-generation information technologies.

Among 2D vdW magnets, CrSBr stands out as an environmentally stable magnetic semiconductor that combines ferromagnetic interaction within each layer with antiferromagnetic correlation between adjacent layers \cite{Wilson2021}. Consequently, magnetic order directly reshapes both the underlying electronic structure and the excitonic response \cite{klein2022,Wilson2021,komar2024,smiertka2025unraveling}. Excitons are Coulomb-bound electron-hole pairs that interact strongly with light through their dynamic dipole and thus provide a direct route to optical readout and information transfer \cite{Wang2018RMPExcitonsTMD,Tagarelli2023NatPhotonExcitonTransport,TabatabaVakili2024NatCommunCrSBrExcitonMagnetism,Diederich2025NatNanoExcitonMagnonCrSBr,shao2025,MarquesMoros2023}. In CrSBr, reflectance offers direct access to these magnetic configurations through the excitonic dielectric response, providing a particularly direct optical readout of the underlying layer-resolved magnetic order.

Here, we establish CrSBr as a naturally reconfigurable magneto-optical metamaterial that enables deterministic optical readout of layer-resolved magnetic domain structures and their switching through metastable intermediate states. Applied magnetic fields and interfacial coupling provide deterministic routes to stabilize, suppress, or reshape optically readable intermediate magnetic states, as demonstrated here interfacing CrSBr with the Néel-type antiferromagnetic vdW material MnPS$_3$.

These combined properties make CrSBr promising for spin-optoelectronics and, more broadly, position it as a platform for intelligent matter with clear potential for neuromorphic functionality.

\section{Results}

\begin{figure*}[t]
\centering
\includegraphics[width=\textwidth]{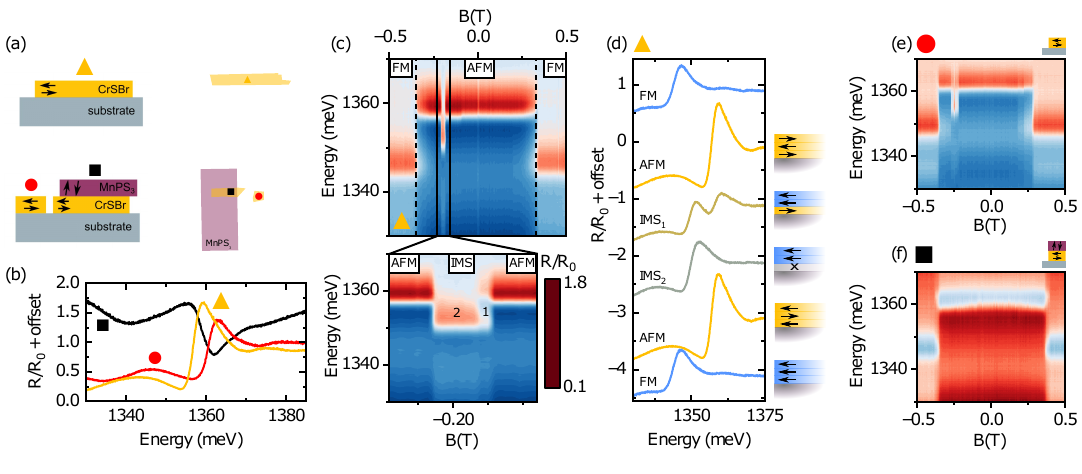}
\caption{\textbf{Intermediate magnetic states in 3L CrSBr and their modification by interfacial coupling to MnPS$_3$.}
\textbf{(a)} Optical micrographs and schematics of the investigated 3L CrSBr regions before and after MnPS$_3$ capping. During the transfer of the MnPS$_3$ flake, the underlying CrSBr flake tore, resulting in MnPS$_3$-capped and uncapped, spatially separated CrSBr regions. The square, triangle, and circle denote the corresponding regions/configurations and are used consistently throughout all panels of the figure. 
\textbf{(b)} Representative reflectance spectra $R/R_0$  measured at zero magnetic field from the marked regions. 
\textbf{(c)} Field-dependent evolution of the reflectance spectra $R/R_0$ of the 3L flake before capping.
\textbf{(d)} Representative spectra of distinct magnetic signatures recorded at selected external magnetic fields. In bare 3L CrSBr, the intermediate magnetic state (iMS) is identified by a spectrally distinct response between the AFM and FM limits. Two characteristic spectra are resolved within the intermediate-state field window, indicating a nontrivial optical response of the trilayer in the iMS regime.
\textbf{(e)} Field-dependent evolution of the reflectance spectra $R/R_0$ of the 3L flake after MnPS$_3$ capping, recorded in the uncapped, separated region (upper panel, red circle) and in the capped region (lower panel, black square). After capping, the iMS signatures disappear and no distinct intermediate-state window remains, indicating a strong interfacial modification of the trilayer magnetic response.}
\label{fig1}
\end{figure*}

\begin{figure}[t]
\centering
\includegraphics{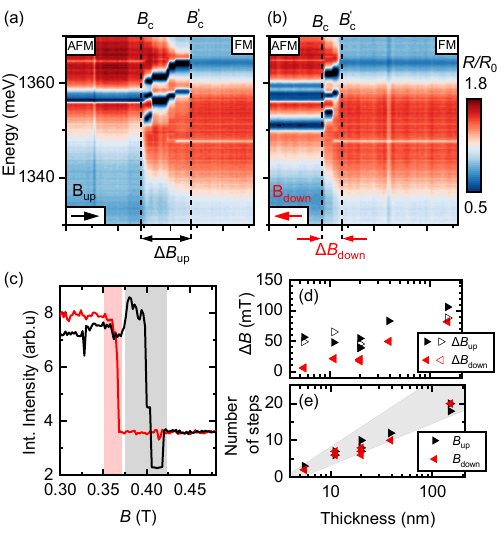}
\caption{\textbf{Thickness dependence of field-driven optical switching in CrSBr.} \textbf{(a,b)} Magneto-reflectance spectra $R/R_0$ with $B$ applied along the easy $b$-axis for a 20\,nm thick CrSBr flake. The reflectance ratio $R/R_0$ is color-coded. Magnetic-field up- and down-sweeps, $B_\mathrm{up}$ and $B_\mathrm{down}$, are shown. Vertical dashed lines separate the AFM, iMS, and FM regimes and indicate the critical switching fields $B_\mathrm{c}$ and $B'_\mathrm{c}$. 
\textbf{(c)} Integrated $R/R_0$ reflectance ratio in the AFM exciton range (1364.1--1365.3\,meV) for the 20\,nm flake, shown for up- and down-sweeps. The shaded regions mark the iMS windows. 
\textbf{(d)} Magnetic-field window $\Delta B$ of the iMS phase as a function of layer thickness for up- and down-sweeps. 
\textbf{(e)} Number of steps in the iMS regime as a function of layer thickness for up- and down-sweeps. Full symbols stem from reflectance and open symbols from PL measurements. Panels (d,e) summarize the systematic thickness dependence of the intermediate-state regime. }
\label{fig2}
\end{figure}

\begin{figure}[t]
\centering
\includegraphics{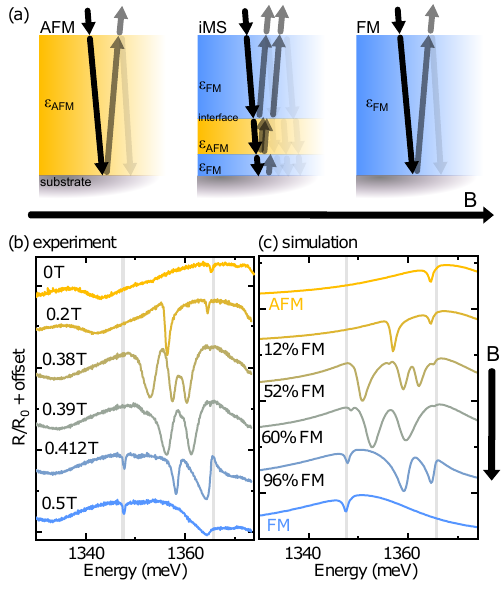}
\caption{\textbf{Comparison of experimental and simulated spectra for 25-layer CrSBr.}
\textbf{(a)} Schematic dielectric profiles for AFM, iMS, and FM configurations, illustrating alternating AFM and FM layers in the magnetic multilayer stack. 
\textbf{(b)} Experimental magneto-reflectance spectra $R/R_0$ for selected magnetic-field values with the field applied along the CrSBr easy axis, where 0\,T and 0.5\,T correspond to AFM and FM order, respectively, and intermediate fields correspond to iMS configurations; traces are offset for clarity. 
\textbf{(c)} Simulated magneto-reflectance spectra obtained from the TMM for AFM, FM, and representative iMS configurations with increasing FM content (given in \%). Simulations are based solely on $\varepsilon_\mathrm{AFM}(\omega)$, $\varepsilon_\mathrm{FM}(\omega)$, and the layer thickness, and are not fitted to the experimental data.}
\label{fig3}
\end{figure}

The reconfigurable magnetic states of CrSBr are studied systematically by magneto-reflectance measurements on thin CrSBr films interfaced with collinear N\'eel-ordered antiferromagnetic MnPS$_3$ (Fig.~\ref{fig1}) and for various CrSBr thicknesses (Fig.~\ref{fig1}-\ref{fig4}). Unless noted otherwise, all measurements are performed with the external magnetic field applied along the in-plane easy axis of CrSBr, i.e. the crystallographic $b$ direction, at a cold-finger temperature of approximately 4\,K. The heterostructures are prepared by micromechanical exfoliation and dry transfer onto standard Si/SiO$_2$ substrates with a 100\,nm oxide thickness.

Figure \ref{fig1} introduces the investigated trilayer regions before and after MnPS$_3$ capping and defines the symbols used to label the corresponding sample areas and magnetic configurations. It compares the optical and magneto-optical response of the CrSBr flake before and after MnPS$_3$ capping.

In the trilayer limit, the magneto-reflectance data resolve the characteristic intermediate-state (iMS) regime that appears between the AFM and FM configurations [Fig.~\ref{fig1}(c)], consistent with previous reports on 3L CrSBr \cite{TabatabaVakili2024NatCommunCrSBrExcitonMagnetism}. Figure \ref{fig1}(d) shows representative spectra of the AFM, FM, and intermediate-state responses observed along the transition. A closer inspection reveals that the 3L intermediate-state window contains two distinct non-additive optical signatures, denoted iMS$_1$ and iMS$_2$, which cannot be explained by simple AFM/FM averaging. In particular, iMS$_2$ is characterized by a single spectral feature at an energy between the AFM and FM limits. While iMS$_1$ may be viewed as a superposition of an AFM-like contribution and iMS$_2$, the very occurrence of iMS$_2$ is nontrivial and cannot be accounted for by simple AFM/FM averaging. The existence of these two distinct intermediate spectra points to a discrete step-by-step switching process, which is captured below by an optical multilayer model in terms of nearly layer-resolved magnetic reconfiguration.

The intermediate-state response is strongly modified by MnPS$_3$ capping, revealing the sensitivity of the trilayer iMS to interfacial coupling. A comparison of the magneto-reflectance spectra of bare and MnPS$_3$-covered CrSBr [Fig.~\ref{fig1}(e)] shows that the layer-by-layer reorientation is significantly altered by the magnetic interface. Bare 3L CrSBr exhibits a hysteretic AFM-to-FM transition with a well-resolved intermediate-state region. In the capped region, by contrast, the characteristic iMS signatures are no longer resolved and the distinct intermediate-state window disappears, indicating that interfacial coupling suppresses the iMS response observed in bare 3L CrSBr. This behaviour is consistent with interfacial pinning and proximity-induced modifications of the magnetic energy landscape. Thus, while the existence of an iMS in 3L CrSBr is known, our measurements show that interfacial coupling can substantially reshape and, in this case, fully suppress the optically resolved intermediate-state response. The trilayer therefore represents the ultrathin limit of the broader behaviour developed below for thicker flakes, where discrete layer-resolved AFM/FM configurations give rise to optically resolvable intermediate states.

For thicker flakes, we focus on uncapped CrSBr with seven layers or more, where the intermediate-state regime evolves into a richer, thickness-dependent cascade of configurations. Figure \ref{fig2} summarizes this behaviour using a representative 20\,nm flake together with the systematic thickness dependence of the number of switching steps and the iMS field window.

A few discrete jumps appear in the magneto-reflectance spectra [Fig.~\ref{fig2}(a,b)] for a 20\,nm thick CrSBr flake consisting of 25 layers, assuming a monolayer thickness of $d_{\mathrm{ML}} \approx 0.78$\,nm \cite{lee2021magnetic,pellet2025lateral}. Importantly, these jumps are observed for both magnetic-field up- (a) and down-sweeps (b), but their number, position, and spectral evolution differ markedly between the two sweep directions. Both the energy and the reflectance ratio at a given energy are highly sensitive to the different iMS states. The integrated reflectance ratio in the energy range of the exciton in the AFM state for the 20\,nm CrSBr flake is shown in Figure \ref{fig2}(c) for magnetic-field up- (black) and down-sweeps (red). The iMS window is marked for both traces by the shaded area. While the overall iMS signatures emerge in both up- and down-sweeps, the integrated reflectance ratio $R/R_0$, the number of steps, and the magnetic-field window $\Delta B$ depend strongly on the sweep direction and hence on the path of the magnetic phase transition.  

This sweep-direction dependence is also reflected in the critical fields $B_\mathrm{c}$ and $B'_\mathrm{c}$, which exhibit the well-known hysteresis for switching along the magnetic easy axis \cite{StonerWohlfarth1948,Dieny2017RMP_PMA,TabatabaVakili2024NatCommunCrSBrExcitonMagnetism,BoixConstant2025AdvMaterCrSBrHysteresis}. The critical fields separate the iMS region from AFM and FM configurations, respectively, and define the iMS window through $\Delta B = B'_\mathrm{c} - B_\mathrm{c}$. The iMS region reveals numerous intermediate states and pronounced hysteresis between up- and down-sweeps, resulting in a substantially wider magnetic-field window for the up-sweep than for the down-sweep, $\Delta B_\mathrm{up} > \Delta B_\mathrm{down}$. The reduced $\Delta B$ and lower critical fields for the down-sweep are consistent with the hysteretic behaviour of magnetization along the easy axis. 

By comparing samples of different thicknesses, we find that the observed magnetic-field-dependent trends are universal and become more pronounced with increasing layer number and hence thickness, as summarized in Figure \ref{fig2}(d,e) and Supplementary Fig.~3. In particular, both the number of steps in the iMS regime and the magnetic-field window $\Delta B$ increase with thickness. These systematic trends point to an increasing multiplicity and stability of metastable intermediate configurations with thickness. 

The evolution of the spectra in the iMS configuration is consistent with field-driven multi-step, layer-resolved switching between AFM and FM order \cite{Wilson2021,alapatt2025highly}. While previous magneto-optical studies described the transition in terms of gradual conversion toward the FM state and coexistence of AFM- and FM-like spectral signatures \cite{krelle2025magnetic}, here we show that the intermediate spectra can be understood as fingerprints of discrete metastable layered AFM/FM configurations. In this scenario, the homogeneous AFM or FM vdW stack transforms into a multilayered out-of-plane magnetic-domain system with AFM and FM sub-stacks of variable thickness.

A representative multilayer domain configuration is schematically shown in Figure \ref{fig3}(a). We emphasize, however, that many additional configurations with multiple AFM/FM interfaces are possible. In the iMS regime, the layered magnetic domain structure is directly linked to an optical multilayer system and can be viewed as a reconfigurable layered optical metamaterial. As a result, different magnetic configurations give rise to distinct optical fingerprints, which provide direct access to the layer-by-layer spin reconfiguration and allow one to trace the controlled manipulation of layered magnetic domains and the emergence of metastable reconfigurable states.

The microscopic origin of the sensitivity of the optical spectra to the long-range magnetic order and hence the out-of-plane domain structure in the iMS is the magnetization-dependent spatial extent of the relevant electronic states. In antiferromagnetic alignment, the electronic wave-functions, and thus the exciton, are confined to a single monolayer due to suppressed interlayer hopping because of opposite spin-orientation in adjacent layers. The dielectric response of the exciton is therefore monolayer-like, and the exciton energy is higher due to the higher single-particle energy \cite{heissenbuttel2025quadratic}. In the ferromagnetic configuration, parallel spin alignment permits interlayer tunneling, delocalizing the electronic and hence excitonic states across \(\sim 5\) layers and lowering their energies \cite{heissenbuttel2025quadratic}. For ferromagnetic stacks thinner than these 5 layers, the exciton energy is expected to lie between the monolayer-like AFM and bulk-FM limits due to quantum confinement. This magnetization dependence of the excitonic resonance translates directly into a change of the complex dielectric function, i.e.\ $\varepsilon_\mathrm{AFM}(\omega) \neq \varepsilon_\mathrm{FM}(\omega)$, and thereby modifies the propagation and interference of light in the multilayer stack. 

In multilayer CrSBr flakes such as the 20\,nm sample shown in Figure \ref{fig2}, which consists of 25 individual layers, the iMS regime is particularly rich. An analogous behaviour is observed for thicker flakes as well (see Supplementary Information). Successive subsets of layers can flip nearly independently from AFM to FM or vice versa, giving rise to many possible layered magnetic-domain configurations and turning the system into an optical analogue of an artificial superlattice \cite{Yablonovitch1987PRL_PBG, Fink1998Science_DBR, Dufferwiel2015NatCommun_vdWCavity,Sciesiek2020CommsMaterialsCoupledCavities, Zhao2023NatCommun_vdWSuperlattice}.

With a direct bandgap of around 1.3 eV and an unusually high refractive index near its excitonic resonances, even tens-of-nanometers-thick CrSBr flakes exhibit pronounced interference effects dictated by van der Waals stacking rather than external nanofabrication \cite{dirnberger2023magneto,wang2023,diederich2023}. Previous studies have identified several excitonic resonances in CrSBr, including energetically close-lying bulk excitons and a surface-confined state \cite{shao2025}, although their microscopic origin remains under active discussion. These resonances are sharp and carry strong oscillator strength, which leads to efficient coupling to light. As a result, the optical spectra of multilayer CrSBr are shaped not only by the intrinsic excitonic response, but also by light-guiding and interference within the layered crystal. In thicker stacks, this interplay can even lead to hybridized modes commonly described within a polaritonic framework \cite{dirnberger2023magneto}. Excitonic resonances in multilayer CrSBr are therefore not observed in isolation, but in a form modified by the dielectric environment and optical interference within the stack.

The optical response becomes particularly sensitive to magnetic order when individual layers switch between AFM and FM alignment. In the spectral range of interest, the dielectric response of CrSBr is dominated by a single extremely strong and sharp resonance that shifts by about 20\,meV between the AFM and FM states. An external magnetic field applied along the easy axis therefore does not merely shift an excitonic resonance, but modifies $\varepsilon(\omega)$ and thereby the light-guiding properties of the magnetic layer stack.

The modified dielectric response reconfigures the propagation and interference of light in the multilayer stack. The number of interfaces, together with the AFM/FM layer sequence in the iMS, determines the reflectance and PL response (see SI, Figure S5), giving rise to several distinct resonances. This layer sequence depends on magnetic field, sweep direction, and thickness. In other words, the cascade of abrupt changes in the optical spectra corresponds to a discrete reconfiguration of the layered magnetic domain structure. The optical response therefore provides a direct fingerprint of the magnetic configuration. Moreover, the reflectance measurements serve as a sensitive readout of the field-written iMS state, allowing one to infer whether switching starts from the top, bottom, or inner part of the stack.

Transfer-matrix method (TMM) simulations validate this interpretation by showing that the abrupt changes in energy and intensity arise from modified light propagation and interference within the magnetic-field-induced layered domain structure. In Figure \ref{fig3}, measured (b) and simulated (c) reflectance spectra are compared side-by-side for a 25L-CrSBr stack. We compare spectra for a sequence of field-written configurations, starting from the AFM state, passing through representative iMS configurations with increasing FM fraction, and ending in the FM state. The simulations reproduce the key experimental features.

The TMM simulations are based on a minimal model using a single Lorentzian oscillator in $\epsilon(\omega)$, representing the main excitonic resonance, whose frequency redshifts abruptly when the system transitions from AFM to FM order, in agreement with experiment and theory \cite{Wilson2021, heissenbuttel2025quadratic}. We neglected the contribution of high-lying excitons to the dielectric function $\epsilon(\omega)$ because they lie far outside the investigated spectral range and preliminary tests showed that their contribution is negligible in the TMM. Accordingly, we restrict the model to a single well-established excitonic resonance in $\varepsilon(\omega)$ within the spectral range of interest. We find that between the pure AFM and FM situation in the stack, the simulated spectra evolve from a single mode into a ladder of modes. 

The magnetic-field-driven rearrangement of the layered magnetic domains is accompanied by abrupt changes in the optical modes due to altered light propagation and interference conditions. Upon approaching either the fully AFM or fully FM state, the rich spectrum collapses back to a single feature. We deliberately choose not to fit complete, parameterized dielectric functions to the data, as the large number of available parameters allows a plethora of possible solutions. Instead, our approach makes clear that just the interference effects between layers with $\epsilon_{AFM}$ and $\epsilon_{FM}$ are sufficient to explain the main experimental features, without having to invoke additional excitonic species.

The grey vertical lines shown in the experimental and simulated spectra in Figures \ref{fig3}(b,c) mark the exciton energies in the AFM and FM phases, making it explicit that the observed mode shifts follow the exciton's magnetic response.

\begin{figure}[t]
\centering
\includegraphics{Figure_3_2.pdf}
\caption{\textbf{Magneto-reflectance investigation for 7-layer CrSBr as the lower limit for layered-magnetic domain formation.} Magneto-reflectance spectra \(R/R_0\) for a 7L-CrSBr flake; \textbf{(a)} up-sweep, \(B_\mathrm{up}\), \textbf{(b)} down-sweep, \(B_\mathrm{down}\). Vertical dashed lines indicate critical switching fields.
\textbf{(c)} Selected reflectance spectra for AFM, iMS and FM state for up-sweep. The asterisk (*) marks an additional resonance that appears only within the intermediate-state window for \(B_\mathrm{up}\). \textbf{(d)} Simulated magneto-reflectance spectra using the TMM for pure AFM, pure FM and iMS (4L FM, 1L AFM, 2L FM) configuration depicted at the right. }\label{fig4}
\end{figure}

We now address the ultrathin limit of CrSBr, where the intermediate-state regime approaches its lower thickness boundary. We illustrate this regime using a 7-layer CrSBr film with a thickness of approximately 5.5\,nm, as determined by atomic force microscopy. As shown in Fig.~\ref{fig4}(a,b), the magneto-reflectance spectra evolve very differently for up- (a) and down-sweeps (b) in this ultrathin regime. During the up-sweep, the system exhibits the established transition from AFM through iMS to FM order, with two critical fields, $B_\mathrm{c}$ and $B'_\mathrm{c}$, separating distinct spectral regimes. In contrast, the down-sweep reveals only a single critical field, $B_\mathrm{c}=B'_\mathrm{c}$, consistent with a direct transition from FM to AFM. In other words, the up-sweep is consistent with layer-by-layer switching through an iMS, whereas the down-sweep points to a direct transition without layered domain formation. Representative experimental reflectance spectra for AFM, iMS and FM states are shown in Figure \ref{fig4}(c). In the iMS range of the up-sweep, we observe the coexistence of spectral signatures belonging to excitons in both antiferromagnetic (AFM) and ferromagnetic (FM) configurations, as well as an additional resonance located energetically between them. 

The additional mode, marked by an asterisk in Figure \ref{fig4}(c), can only be reproduced in the TMM by introducing an additional intermediate resonance; for thicker films, by contrast, the experimental spectra are well captured using only AFM and FM layers. The required intermediate resonance must therefore lie energetically between the AFM and FM excitons. This is consistent with quantum confinement, which increases the exciton energy of FM regions thinner than 5\,layers and drives it toward the AFM limit in the monolayer case. In thicker films, FM domains thinner than about five layers are expected to be energetically unfavorable. In the few-layer limit, however, this thickness dependence can no longer be neglected. Accordingly, a metastable iMS configuration in 7L CrSBr consisting of two FM sub-stacks separated by a single AFM layer, as sketched in Figure \ref{fig4}(d), can account for the observed spectra. Other microscopic origins, such as additional excitonic species, cannot be fully excluded, but they neither explain the reproducible absence of this additional resonance in the magnetic-field down-sweep nor are they supported by existing theory \cite{heissenbuttel2025quadratic}. The asymmetry between up- and down-sweeps may additionally be linked to reduced exciton localization in the FM state, which may favor a direct FM-to-AFM transition in the 7L limit.

This thin-layer confinement picture may also provide a natural framework for understanding the trilayer iMS spectra in Figure \ref{fig1}. In particular, the two distinct spectral signatures observed in the 3L intermediate-state window are consistent with different layer-resolved configurations: one shows two features, compatible with a 2L+1L-like response, whereas the other is dominated by a single intermediate-energy feature, consistent with a 2L-like optical response and a third, effectively silent layer. Here, ``silent'' does not imply the absence of magnetic order, but rather that the corresponding layer remains magnetically active while not giving rise to a distinct separately resolved optical feature in the measured reflectance. Within the TMM, this phenomenology is captured by an effective 2L-like optical response, i.e. the spectrum is well reproduced without an additional separately resolved optical contribution from the third layer. Such behavior can arise because, in the thin-layer limit, the exciton energy depends sensitively on both the effective magnetic thickness and the local magnetic configuration \cite{Wilson2021,heissenbuttel2025quadratic}. This interpretation is consistent with the existence of magnetically nonuniform configurations and phase boundaries in atomically thin CrSBr \cite{Tschudin2024,pellet2025lateral}.

The optical response of CrSBr under an external magnetic field applied along the easy axis can be understood within a unified magnetic multilayer picture. Field-driven layer-by-layer switching generates metastable AFM/FM stacks with distinct optical fingerprints governed by the magnetization-dependent dielectric response. Within this framework, the TMM reproduces the magneto-reflectance spectra across different thicknesses without invoking additional excitonic species, while the ultrathin limit reveals that quantum confinement must be taken into account for very small FM sub-stacks. This establishes optical reflectance as a direct probe of layer-resolved magnetic reconfiguration in CrSBr.

The resulting tunability of the optical response makes CrSBr a naturally reconfigurable magneto-optical platform in which metastable magnetic configurations can be accessed and read out optically. Beyond magnetic-field control, interfacial coupling and other external stimuli could offer additional routes for manipulating these states. These properties make CrSBr promising for spin-optoelectronics and, more broadly, position it as a platform that fulfills key requirements for intelligent matter and offers clear potential for neuromorphic functionality.

\section{Methods}\label{sec11}
\subsection{Sample preparation.}
CrSBr flakes were mechanically exfoliated from bulk crystals onto Si substrates with a 100-nm SiO$_2$ layer. Flake thicknesses were verified by atomic force microscopy (AFM). In the 3L case, the CrSBr flake was capped with a thick MnPS$_3$ flake that was mechanically exfoliated and dry-transferred.

\subsection{Optical measurements.}
All experiments were performed at 4\,K in a closed-cycle cryostat equipped with optical access and superconducting magnets. The detection polarization axis was aligned along the crystallographic $b$-axis of CrSBr, which exhibits a strong optical response, while the orthogonal axis ($a$) yields negligible signal. PL was excited with a continuous-wave 532-nm laser. Reflectance spectra were acquired using a laser-driven broadband white-light source. Both PL and reflectance were collected in a backscattering geometry with linear polarization selection and dispersed in a spectrometer.

\subsection{Magnetic field configuration.}
Magnetic fields were applied along the easy axis of CrSBr ($b$-axis), with values up to 0.5\,T. The step size in magnetic field was 0.002\,T. For comparison, measurements were also performed with the field along the intermediate a-axis, up to 3\,T, confirming the expected anisotropy; these data are provided in the Supplementary Information.

\subsection{Data analysis.}
Spectra were background-corrected and normalized to a reference signal from the bare substrate.

\subsection{Modeling.} The multilayer optical response was simulated with the transfer-matrix method (TMM). AFM- and FM-aligned layers were assigned distinct dielectric functions, \(\varepsilon_{\mathrm{AFM}}(\omega)\) and \(\varepsilon_{\mathrm{FM}}(\omega)\), each containing a single Lorentz oscillator; background indices and oscillator strengths were fixed to the same value. The only free parameters were the resonance energies in the AFM and FM cases, which were extracted by fitting the low-field (AFM) and high-field (FM) reflectance spectra of the thick flake, respectively. To capture the few-layer regime, we introduced an explicit thickness dependence of the FM exciton energy, \(E_X^{\mathrm{FM}}(N)\), which decreases with the ferromagnetic slab thickness \(N\) and saturates for \(N\gtrsim5\) monolayers (bulk-like limit). The exact values for \(E_X^{\mathrm{FM}}(N)\) and other parameters used for calculations are given in the supplementary information, along with a comparison to a simplified model in which the exciton energy is taken to be thickness-independent. Both reproduce the emergence of multiple modes and their abrupt field-induced rearrangements.

We did not model the magnetization dynamics. Instead, we assumed that the ferromagnetic phase fraction in the stack increases monotonically with the applied field.

\backmatter

\bmhead{Supplementary Information}
See Supplementary Information for additional methods, figures, and tables.

\bmhead{Acknowledgements}

The authors gratefully acknowledge financial support by the German Science Foundation (DFG) via Grants No.443274199 and 556436549 (WU 637/7-2,8-1), and the priority program 2244 (2DMP) through start-up funding. This project has received funding from the European Union's Horizon Europe research and innovation program under grant agreement No 101130224 'JOSEPHINE'. 
T.D. acknowledges financial support from the Deutsche Forschungsgemeinschaft (DFG, German Research Foundation) through Project 
No. 426726249 (DE 2749/2-1 and DE 2749/2-2).
The authors gratefully acknowledge the Gauss Centre for Supercomputing e.V. (www.gauss-centre.eu) for funding this project by providing  computing time through the John von Neumann Institute for Computing (NIC) on the GCS Supercomputer JUWELS at J{\"u}lich Supercomputing Centre 
(JSC) \cite{JUWELS}.
Z.S. was supported by project LUAUS25268 from the Ministry of Education, Youth and Sports (MEYS) and by the project Advanced Functional Nanorobots (reg. No. CZ.02.1.01/0.0/0.0/15\_003/0000444 financed by the EFRR) and by the ERC-CZ program (project LL2101) from Ministry of Education, Youth and Sports (MEYS).
Z.S. acknowledges the assistance provided by the Advanced Multiscale Materials for Key Enabling Technologies project, supported by the Ministry of Education, Youth, and Sports of the Czech Republic. Project No. CZ.02.01.01/00/22\_008/0004558, Co-funded by the European Union.


\clearpage
\setcounter{figure}{0}
\renewcommand{\thefigure}{S\arabic{figure}}
\makeatletter
\renewcommand{\theHfigure}{S\arabic{figure}} 
\makeatother

\begin{center}
\textbf{\large Supplementary Information:\\
Optical Readout of Reconfigurable Layered Magnetic Domain Structure in CrSBr}
\end{center}

\begin{figure}[h]
\centering
\includegraphics{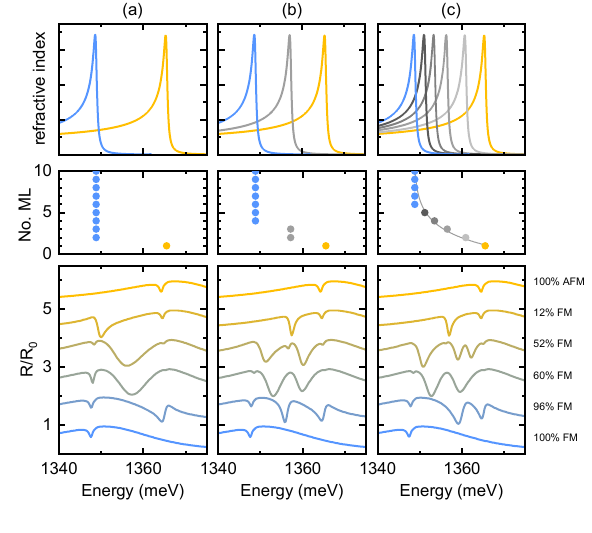}
\caption{Comparison between three different approaches to the construction of the dielectric function used in the calculations, along with the subsequent result of the model. In each column: the top row shows the dispersion of the refractive index $n(E)$; the middle row shows the energy of the excitonic resonance depending on the number of layers, all other than energy parameters of the excitonic resonance used to the construction of the dielectric function are the same. The bottom row shows the normalized reflectance $R/R_0$ for exemplary configurations of AFM/FM stacks with different FM content (given in \%). First approach (a) is differing only the monolayer exciton energy (AFM state) and FM state (the number of layers $\geqslant 2$). Even with this simple model we can observe characteristics of the intermediate AFM/FM stack --- occurrence of more features in the spectrum and blueshifts. (b) The second approach with additional intermediate energy for thin layers (2--3L); The approach presented in (c) was used in the main paper.}\label{sup:fig1}
\end{figure}

\begin{figure}
\centering
\includegraphics{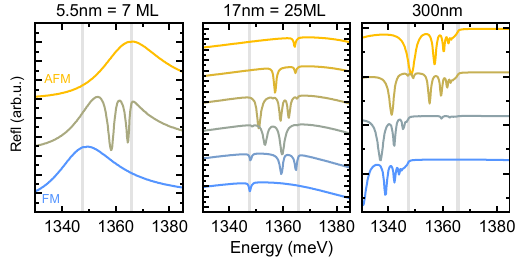}
\caption{Left: calculated reflectance spectra for flakes of 7\,ML, 25\,ML, and 300\,ML thickness. Spectra are shown for the pure AFM (top), FM (bottom) and intermediate states between. For 7\,ML, the sequence corresponds to one AFM layer sandwiched between two ferromagnetic layers with thickness of 4 (top) and 2 (bottom). For 25\,ML and 300\,nm, representative mixed configurations are presented. The content of the FM layers is increasing from top (AFM) to bottom (FM) of the graphs.}\label{sup:fig2}
\end{figure}

\begin{figure}
\centering
\includegraphics[height=0.75\textheight, keepaspectratio]{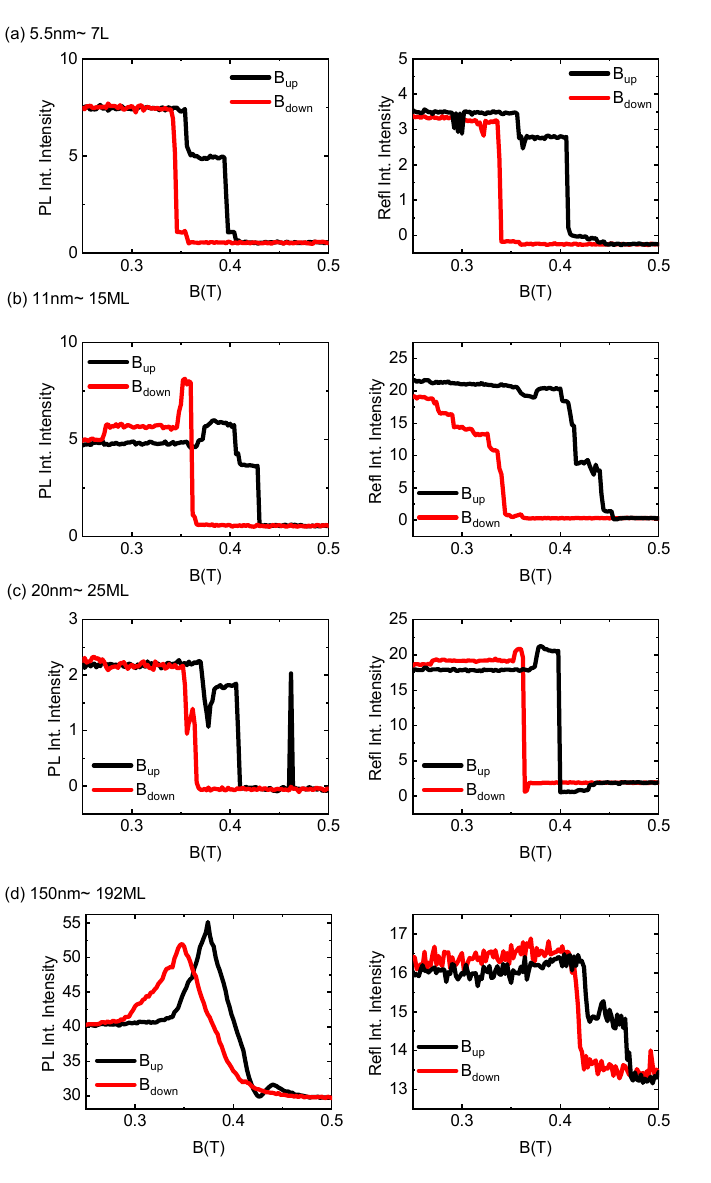}
\caption{Integrated signal for the AFM exciton range vs.\ applied magnetic field of different direction (increasing -- black and decreasing -- red) for different thicknesses of the flakes for photoluminescence and reflectance scans. Such integrated intensity can give information about number of switches in the intermediate magnetic state region and the range of the magnetic field of such.}\label{sup:fig5}
\end{figure}

\begin{figure}
\centering
\includegraphics[width=\textwidth]{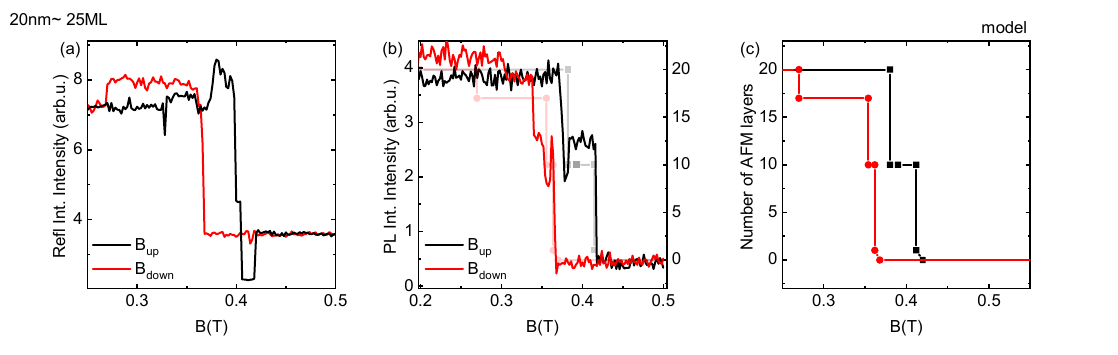}
\caption{Integrated signal for the AFM exciton range vs.\ applied magnetic field of different direction (increasing -- black and decreasing -- red) for a 20\,nm flake: reflectance (a) and photoluminescence (b). Underneath measured data in faint lines is presented the graph (c) which represents the number of AFM layers obtained from the model. That suggests that in the calibrated system the intermediate state can be read out without direct fitting.}\label{sup:fig6}
\end{figure}

\begin{figure}
\centering
\includegraphics[width=\textwidth]{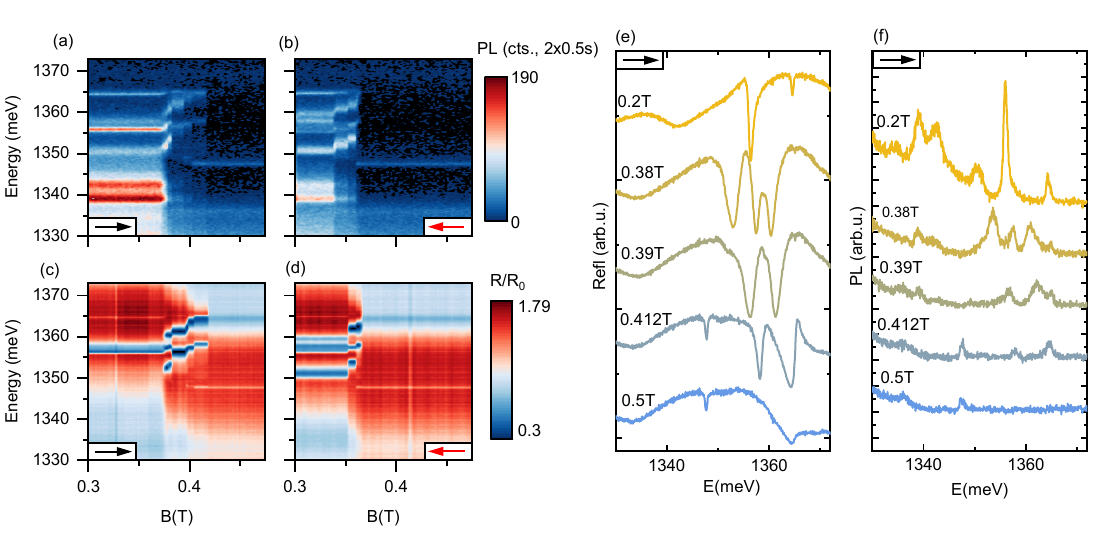}
\caption{Comparison between photoluminescence (a,b) and reflectance (c,d) scans for a 20\,nm sample for increasing (left) and decreasing (right) external magnetic field. On the right-hand side panels exemplary spectra taken from the increasing scans are presented both for reflectance (e) and photoluminescence (f). Scans in the same direction of the magnetic field give complementary PL and reflectance spectra.}\label{sup:fig3}
\end{figure}

\begin{figure}
\centering
\includegraphics{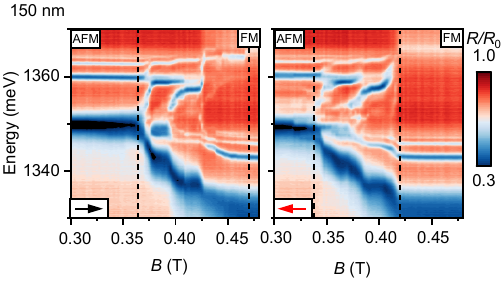}
\caption{Magneto-reflectance spectra $R/R_0$ of a 150\,nm thick CrSBr flake as a function of magnetic field $B$ applied along the easy $b$-axis and photon energy. Left and right panels show the field up- and down-sweeps, respectively, as indicated by the arrows. Vertical dashed lines mark the boundaries between the AFM, intermediate-state, and FM regimes. The maps reveal a pronounced hysteretic evolution with multiple intermediate spectral rearrangements between the AFM and FM limits.}\label{sup:fig7}
\end{figure}


\begin{thebibliography}{41}

\bibitem{Jungwirth2016} Tomas Jungwirth, Xavier Marti, Peter Wadley, and Jairo Wunderlich, Antiferromagnetic spintronics. \textit{Nature Nanotechnology} \textbf{11} (3), 231--241 (2016). DOI: \href{https://doi.org/10.1038/nnano.2016.18}{10.1038/nnano.2016.18}.

\bibitem{Baltz2018} V. Baltz, A. Manchon, M. Tsoi, T. Moriyama, T. Ono, and Y. Tserkovnyak, Antiferromagnetic spintronics. \textit{Reviews of Modern Physics} \textbf{90} (1), 015005 (2018). DOI: \href{https://doi.org/10.1103/RevModPhys.90.015005}{10.1103/RevModPhys.90.015005}.

\bibitem{Zutic2004} I. {{\v Z}uti{\'c}}, J. Fabian, and S. Das Sarma, Spintronics: Fundamentals and applications. \textit{Reviews of Modern Physics} \textbf{76} (2), 323--410 (2004). DOI: \href{https://doi.org/10.1103/RevModPhys.76.323}{10.1103/RevModPhys.76.323}.

\bibitem{kaspar2021rise} Corinna Kaspar, Bart Jan Ravoo, Wilfred G van der Wiel, Seraphine V Wegner, and Wolfram HP Pernice, The rise of intelligent matter. \textit{Nature} \textbf{594} (7863), 345--355 (2021). DOI: \href{https://doi.org/10.1038/s41586-021-03453-y}{10.1038/s41586-021-03453-y}.

\bibitem{Huang2017} Bevin Huang, Genevieve Clark, Efren Navarro-Moratalla, Dahlia R. Klein, Ran Cheng, Kyle L. Seyler, Ding Zhong, Emma Schmidgall, Michael A. McGuire, David H. Cobden, Wang Yao, Di Xiao, Pablo Jarillo-Herrero, and Xiaodong Xu, Layer-dependent ferromagnetism in a van der {W}aals crystal down to the monolayer limit. \textit{Nature} \textbf{546}, 270--273 (2017). DOI: \href{https://doi.org/10.1038/nature22391}{10.1038/nature22391}.

\bibitem{Gong2017} Cheng Gong, Lin Li, Zhenglu Li, Huiwen Ji, Alex Stern, Yang Xia, Ting Cao, Wei Bao, Chenzhe Wang, Yuan Wang, Zhiqiang Q. Qiu, Robert J. Cava, Steven G. Louie, Jing Xia, and Xiang Zhang, Discovery of intrinsic ferromagnetism in two-dimensional van der {W}aals crystals. \textit{Nature} \textbf{546}, 265--269 (2017). DOI: \href{https://doi.org/10.1038/nature22060}{10.1038/nature22060}.

\bibitem{Gong2019Science2DMagnets} Cheng Gong and Xiang Zhang, Two-dimensional magnetic crystals and emergent heterostructure devices. \textit{Science} \textbf{363} (6428), eaav4450 (2019). DOI: \href{https://doi.org/10.1126/science.aav4450}{10.1126/science.aav4450}.

\bibitem{Blei2021Review} Mark Blei, Jose L. Lado, Qing Song, Debojyoti Dey, Onur Erten, Victor Pardo, Riccardo Comin, Sefaattin Tongay, and Antia S. Botana, Synthesis, engineering, and theory of {2D} van der {W}aals magnets. \textit{Applied Physics Reviews} \textbf{8} (2), 021301 (2021). DOI: \href{https://doi.org/10.1063/5.0025658}{10.1063/5.0025658}.

\bibitem{TabatabaVakili2024NatCommunCrSBrExcitonMagnetism} Farsane Tabataba-Vakili, Huy P. G. Nguyen, Anna Rupp, Kseniia Mosina, Anastasios Papavasileiou, Kenji Watanabe, Takashi Taniguchi, Patrick Maletinsky, Mikhail M. Glazov, Zdenek Sofer, Anvar S. Baimuratov, and Alexander H{\"o}gele, Doping-control of excitons and magnetism in few-layer {CrSBr}. \textit{Nature Communications} \textbf{15}, 4735 (2024). DOI: \href{https://doi.org/10.1038/s41467-024-49048-9}{10.1038/s41467-024-49048-9}.

\bibitem{Bae2022} Junyoung Bae, Kwanghee Kim, Gwangsu Lee, Sungmin Jang, Takashi Taniguchi, Kenji Watanabe, Je-Geun Kim, Je-Geun Park, Hee Cheul Lee, Changgu Lee, Seungjin Cha, Gyeonghun Yeom, Jin-Hong Park, Hyeonsik Cheong, Young Hee Lee, and Kwang S. Kim, Exciton--magnon coupling in a 2{D} magnet. \textit{Nature} \textbf{609} (7926), 282--286 (2022). DOI: \href{https://doi.org/10.1038/s41586-022-05071-3}{10.1038/s41586-022-05071-3}.

\bibitem{cenker2022reversible} John Cenker, Shivesh Sivakumar, Kaichen Xie, Aaron Miller, Pearl Thijssen, Zhaoyu Liu, Avalon Dismukes, Jordan Fonseca, Eric Anderson, Xiaoyang Zhu, and others, Reversible strain-induced magnetic phase transition in a van der {W}aals magnet. \textit{Nature Nanotechnology} \textbf{17} (3), 256--261 (2022). DOI: \href{https://doi.org/10.1038/s41565-021-01052-6}{10.1038/s41565-021-01052-6}.

\bibitem{deng2018gate} Yujun Deng, Yijun Yu, Yichen Song, Jingzhao Zhang, Nai Zhou Wang, Zeyuan Sun, Yangfan Yi, Yi Zheng Wu, Shiwei Wu, Junyi Zhu, and others, Gate-tunable room-temperature ferromagnetism in two-dimensional {Fe$_3$GeTe$_2$}. \textit{Nature} \textbf{563} (7729), 94--99 (2018). DOI: \href{https://doi.org/10.1038/s41586-018-0626-9}{10.1038/s41586-018-0626-9}.

\bibitem{chen2019direct} Weijiong Chen, Zeyuan Sun, Zhongjie Wang, Lehua Gu, Xiaodong Xu, Shiwei Wu, and Chunlei Gao, Direct observation of van der {W}aals stacking--dependent interlayer magnetism. \textit{Science} \textbf{366} (6468), 983--987 (2019). DOI: \href{https://doi.org/10.1126/science.aav1937}{10.1126/science.aav1937}.

\bibitem{song2021direct} Tiancheng Song, Qi-Chao Sun, Eric Anderson, Chong Wang, Jimin Qian, Takashi Taniguchi, Kenji Watanabe, Michael A McGuire, Rainer St{\"o}hr, Di Xiao, and others, Direct visualization of magnetic domains and moir{\'e} magnetism in twisted {2D} magnets. \textit{Science} \textbf{374} (6571), 1140--1144 (2021). DOI: \href{https://doi.org/10.1126/science.abj7478}{10.1126/science.abj7478}.

\bibitem{Wilson2021} Nathan P Wilson, Kihong Lee, John Cenker, Kaichen Xie, Avalon H Dismukes, Evan J Telford, Jordan Fonseca, Shivesh Sivakumar, Cory Dean, Ting Cao, and others, Interlayer electronic coupling on demand in a 2{D} magnetic semiconductor. \textit{Nature Materials} \textbf{20} (12), 1657--1662 (2021).

\bibitem{klein2022} Julian Klein, Benjamin Pingault, Matthias Florian, Marie-Christin Hei{\ss}enb{\"u}ttel, Alexander Steinhoff, Zhigang Song, Kierstin Torres, Florian Dirnberger, Jonathan B Curtis, Mads Weile, and others, The bulk van der Waals layered magnet CrSBr is a quasi-1D material. \textit{ACS Nano} \textbf{17} (6), 5316--5328 (2023). DOI: \href{https://doi.org/10.1021/acsnano.2c07316}{10.1021/acsnano.2c07316}.

\bibitem{komar2024} R Komar, A {\L}opion, M Goryca, M Rybak, T Wo{\'z}niak, K Mosina, A S{\"o}ll, Z Sofer, W Pacuski, C Faugeras, and others, Colossal magneto-excitonic effects in 2D van der {W}aals magnetic semiconductor {CrSBr}. \textit{arXiv preprint arXiv:2409.00187} (2024). DOI: \href{https://doi.org/10.48550/arXiv.2409.00187}{10.48550/arXiv.2409.00187}.

\bibitem{smiertka2025unraveling} Maciej Smiertka, Michal Rygala, Katarzyna Posmyk, Paulina Peksa, Mateusz Dyksik, Dimitar Pashov, Kseniia Mosina, Zdenek Sofer, Mark van Schilfgaarde, Florian Dirnberger, Michal Baranowski, Swagata Acharya, and Paulina Plochocka, Distinct magneto-optical response of Frenkel and Wannier excitons in CrSBr. \textit{Nature Communications} \textbf{17}, 1777 (2026). DOI: \href{https://doi.org/10.1038/s41467-026-68482-5}{10.1038/s41467-026-68482-5}.

\bibitem{Wang2018RMPExcitonsTMD} Gang Wang, Alexey Chernikov, Mikhail M. Glazov, Tony F. Heinz, Xavier Marie, Thierry Amand, and Benoit Urbaszek, Colloquium: Excitons in atomically thin transition metal dichalcogenides. \textit{Reviews of Modern Physics} \textbf{90} (2), 021001 (2018). DOI: \href{https://doi.org/10.1103/RevModPhys.90.021001}{10.1103/RevModPhys.90.021001}.

\bibitem{Tagarelli2023NatPhotonExcitonTransport} Fedele Tagarelli, Edoardo Lopriore, Daniel Erkensten, Ra{\"u}l Perea-Caus{\'i}n, Samuel Brem, Joakim Hagel, Zhe Sun, Gabriele Pasquale, Kenji Watanabe, Takashi Taniguchi, Ermin Malic, and Andras Kis, Electrical control of hybrid exciton transport in a van der {W}aals heterostructure. \textit{Nature Photonics} \textbf{17}, 615--621 (2023). DOI: \href{https://doi.org/10.1038/s41566-023-01198-w}{10.1038/s41566-023-01198-w}.

\bibitem{Diederich2025NatNanoExcitonMagnonCrSBr} Geoffrey M. Diederich, Mai Nguyen, John Cenker, Jordan Fonseca, Sinabu Pumulo, Youn Jue Bae, Daniel G. Chica, Xavier Roy, Xiaoyang Zhu, Ting Cao, Di Xiao, and Xiaodong Xu, Exciton dressing by extreme nonlinear magnons in a layered semiconductor. \textit{Nature Nanotechnology} \textbf{20} (5), 617--622 (2025). DOI: \href{https://doi.org/10.1038/s41565-025-01890-8}{10.1038/s41565-025-01890-8}.

\bibitem{shao2025} Yinming Shao, Florian Dirnberger, Siyuan Qiu, Swagata Acharya, Sophia Terres, Evan J Telford, Dimitar Pashov, Brian SY Kim, Francesco L Ruta, Daniel G Chica, and others, Magnetically confined surface and bulk excitons in a layered antiferromagnet. \textit{Nature Materials} \textbf{24}, 391--398 (2025). DOI: \href{https://doi.org/10.1038/s41563-025-02129-6}{10.1038/s41563-025-02129-6}.

\bibitem{MarquesMoros2023} Francisco Marques-Moros, Carla Boix-Constant, Samuel Ma{\~n}as-Valero, Josep Canet-Ferrer, and Eugenio Coronado, Interplay between optical emission and magnetism in the van der {W}aals magnetic semiconductor {CrSBr} in the two-dimensional limit. \textit{ACS Nano} \textbf{17}, 13224--13231 (2023). DOI: \href{https://doi.org/10.1021/acsnano.3c00375}{10.1021/acsnano.3c00375}.

\bibitem{lee2021magnetic} Kihong Lee, Avalon H Dismukes, Evan J Telford, Ren A Wiscons, Jue Wang, Xiaodong Xu, Colin Nuckolls, Cory R Dean, Xavier Roy, and Xiaoyang Zhu, Magnetic order and symmetry in the {2D} semiconductor {CrSBr}. \textit{Nano Letters} \textbf{21} (8), 3511--3517 (2021). DOI: \href{https://doi.org/10.1021/acs.nanolett.1c00219}{10.1021/acs.nanolett.1c00219}.

\bibitem{pellet2025lateral} Cl{\'e}ment Pellet-Mary, Debarghya Dutta, M{\"a}rta A Tschudin, Patrick Siegwolf, Boris Gross, David A Broadway, Jordan Cox, Carolin Schrader, Jodok Happacher, Daniel G Chica, and others, Lateral exchange bias for {N}{\'e}el-vector control in atomically thin antiferromagnets. \textit{Nature Communications} \textbf{16} (1), 9725 (2025). DOI: \href{https://doi.org/10.1038/s41467-025-64700-8}{10.1038/s41467-025-64700-8}.

\bibitem{StonerWohlfarth1948} E. C. Stoner and E. P. Wohlfarth, A mechanism of magnetic hysteresis in heterogeneous alloys. \textit{Philosophical Transactions of the Royal Society A} \textbf{240} (826), 599--642 (1948). DOI: \href{https://doi.org/10.1098/rsta.1948.0007}{10.1098/rsta.1948.0007}.

\bibitem{Dieny2017RMP_PMA} B. Dieny and others, Perpendicular magnetic anisotropy at transition metal/oxide interfaces and beyond. \textit{Reviews of Modern Physics} \textbf{89} (2), 025008 (2017). DOI: \href{https://doi.org/10.1103/RevModPhys.89.025008}{10.1103/RevModPhys.89.025008}.

\bibitem{BoixConstant2025AdvMaterCrSBrHysteresis} C. Boix-Constant and others, Programmable Magnetic Hysteresis in Orthogonally-Twisted {CrSBr} Building Blocks. \textit{Advanced Materials} (2025). DOI: \href{https://doi.org/10.1002/adma.202415774}{10.1002/adma.202415774}.

\bibitem{alapatt2025highly} Varghese Alapatt, Francisco Marques-Moros, Carla Boix-Constant, Samuel Manas-Valero, Kirill I. Bolotin, Josep Canet-Ferrer, and Eugenio Coronado, Highly polarized single-photon emission from localized excitons in WSe2/CrSBr heterostructure. \textit{ACS Photonics} \textbf{12}, 3024--3031 (2025). DOI: \href{https://doi.org/10.1021/acsphotonics.5c00144}{10.1021/acsphotonics.5c00144}.

\bibitem{krelle2025magnetic} Lukas Krelle, Ryan Tan, Daria Markina, Priyanka Mondal, Kseniia Mosina, Kevin Hagmann, Regine von Klitzing, Kenji Watanabe, Takashi Taniguchi, Zdenek Sofer, and others, Magnetic correlation spectroscopy in {CrSBr}. \textit{ACS Nano} \textbf{19} (37), 33156--33163 (2025). DOI: \href{https://doi.org/10.1021/acsnano.5c05470}{10.1021/acsnano.5c05470}.

\bibitem{heissenbuttel2025quadratic} Marie-Christin Hei{\ss}enb{\"u}ttel, Pierre-Maurice Piel, Julian Klein, Thorsten Deilmann, Ursula Wurstbauer, and Michael Rohlfing, Quadratic optical response to a magnetic field in the layered magnet {CrSBr}. \textit{Physical Review B} \textbf{111} (7), 075107 (2025). DOI: \href{https://doi.org/10.1103/PhysRevB.111.075107}{10.1103/PhysRevB.111.075107}.

\bibitem{Yablonovitch1987PRL_PBG} Eli Yablonovitch, Inhibited Spontaneous Emission in Solid-State Physics and Electronics. \textit{Physical Review Letters} \textbf{58} (20), 2059--2062 (1987). DOI: \href{https://doi.org/10.1103/PhysRevLett.58.2059}{10.1103/PhysRevLett.58.2059}.

\bibitem{Fink1998Science_DBR} Yoel Fink, Joshua N. Winn, Shanhui Fan, Chiping Chen, Jurgen Michel, John D. Joannopoulos, and Edwin L. Thomas, A dielectric omnidirectional reflector. \textit{Science} \textbf{282} (5394), 1679--1682 (1998). DOI: \href{https://doi.org/10.1126/science.282.5394.1679}{10.1126/science.282.5394.1679}.

\bibitem{Dufferwiel2015NatCommun_vdWCavity} S. Dufferwiel, S. Schwarz, F. Withers, A. A. P. Trichet, F. Li, M. Sich, O. Del Pozo-Zamudio, C. Clark, A. Nalitov, D. D. Solnyshkov, G. Malpuech, K. S. Novoselov, J. M. Smith, M. S. Skolnick, D. N. Krizhanovskii, and A. I. Tartakovskii, Exciton--polaritons in van der {W}aals heterostructures embedded in tunable microcavities. \textit{Nature Communications} \textbf{6}, 8579 (2015). DOI: \href{https://doi.org/10.1038/ncomms9579}{10.1038/ncomms9579}.

\bibitem{Sciesiek2020CommsMaterialsCoupledCavities} Maciej {{\'S}ciesiek}, Krzysztof Sawicki, Wojciech Pacuski, Kamil Sobczak, Tomasz Kazimierczuk, Andrzej Golnik, and Jan Suffczy{\'n}ski, Long-distance coupling and energy transfer between exciton states in magnetically controlled microcavities. \textit{Communications Materials} \textbf{1}, 78 (2020). DOI: \href{https://doi.org/10.1038/s43246-020-00079-x}{10.1038/s43246-020-00079-x}.

\bibitem{Zhao2023NatCommun_vdWSuperlattice} Jiaxin Zhao, Antonio Fieramosca, Kevin Dini, Ruiqi Bao, Wei Du, Rui Su, Yuan Luo, Weijie Zhao, Daniele Sanvitto, Timothy C. H. Liew, and Qihua Xiong, Exciton polariton interactions in van der Waals superlattices at room temperature. \textit{Nature Communications} \textbf{14}, 1512 (2023). DOI: \href{https://doi.org/10.1038/s41467-023-36912-3}{10.1038/s41467-023-36912-3}.

\bibitem{dirnberger2023magneto} Florian Dirnberger, Jiamin Quan, Rezlind Bushati, Geoffrey M Diederich, Matthias Florian, Julian Klein, Kseniia Mosina, Zdenek Sofer, Xiaodong Xu, Akashdeep Kamra, and others, Magneto-optics in a van der {W}aals magnet tuned by self-hybridized polaritons. \textit{Nature} \textbf{620} (7974), 533--537 (2023). DOI: \href{https://doi.org/10.1038/s41586-023-06275-2}{10.1038/s41586-023-06275-2}.

\bibitem{wang2023} Tingting Wang, Dingyang Zhang, Shiqi Yang, Zhongchong Lin, Quan Chen, Jinbo Yang, Qihuang Gong, Zuxin Chen, Yu Ye, and Wenjing Liu, Magnetically-dressed {CrSBr} exciton-polaritons in ultrastrong coupling regime. \textit{Nature Communications} \textbf{14} (1), 5966 (2023). DOI: \href{https://doi.org/10.1038/s41467-023-41688-7}{10.1038/s41467-023-41688-7}.

\bibitem{diederich2023} Geoffrey M Diederich, John Cenker, Yafei Ren, Jordan Fonseca, Daniel G Chica, Youn Jue Bae, Xiaoyang Zhu, Xavier Roy, Ting Cao, Di Xiao, and others, Tunable interaction between excitons and hybridized magnons in a layered semiconductor. \textit{Nature Nanotechnology} \textbf{18} (1), 23--28 (2023). DOI: \href{https://doi.org/10.1038/s41565-022-01259-1}{10.1038/s41565-022-01259-1}.

\bibitem{Tschudin2024} M{\"a}rta A. Tschudin, David A. Broadway, Patrick Siegwolf, Carolin Schrader, Evan J. Telford, Boris Gross, Jordan Cox, Adrien E. E. Dubois, Daniel G. Chica, Ricardo Rama-Eiroa, Elton J. G. Santos, Martino Poggio, Michael E. Ziebel, Cory R. Dean, Xavier Roy, and Patrick Maletinsky, Imaging nanomagnetism and magnetic phase transitions in atomically thin {CrSBr}. \textit{Nature Communications} \textbf{15}, 6005 (2024). DOI: \href{https://doi.org/10.1038/s41467-024-49717-9}{10.1038/s41467-024-49717-9}.

\bibitem{JUWELS} {J\"ulich Supercomputing Centre}, JUWELS Cluster and Booster: Exascale Pathfinder with Modular Supercomputing Architecture at Juelich Supercomputing Centre. \textit{Journal of large-scale research facilities} \textbf{7} (A183) (2021). DOI: \href{https://doi.org/10.17815/jlsrf-7-183}{10.17815/jlsrf-7-183}.

\end{thebibliography}
\end{document}